\documentclass[12pt]{article}
\usepackage{graphicx}
\usepackage{epsfig}
\newcommand{\beq}{\begin{equation}}
\newcommand{\eeq}{\end{equation}}
\newcommand{\beqn}{\begin{eqnarray}}
\newcommand{\eeqn}{\end{eqnarray}}  
\begin{document} 
 
\title{About the quark currents in leptonic decaies of pseudoscalar mesons} 
\author{V.P Efrosinin, A.N. Khotjantsev\\
Institute for Nuclear Research RAS\\
60th October Ave., 7A, 117312, Moscow, Russia}

\date{}
\renewcommand {\baselinestretch} {1.3}

\maketitle
\begin{abstract}

With usage of obvious mechanisms of quark currents the amplitudes of leptonic
decaies of pseudoscalar mesons are investigated. The estimation for a constant
of leptonic decay of the $D^+$ meson is obtained, $f_{D^+} \simeq 0.23~GeV$.  
\end{abstract}

In the paper \cite{Efrosinin:2005mr} by the calculation of form factors in $K_{\mu 3}$
decay in spectator nearing the research of obvious mechanisms of quark
currents was conducted. Thus the output for frameworks of the conventional
phenomenological description of weak decaies of hadrons was carried out.
In the present paper are investigated quark currents in leptonic decaies of
pseudoscalar mesons.

\begin{figure}[ht]
\begin{center}
\epsfig{file=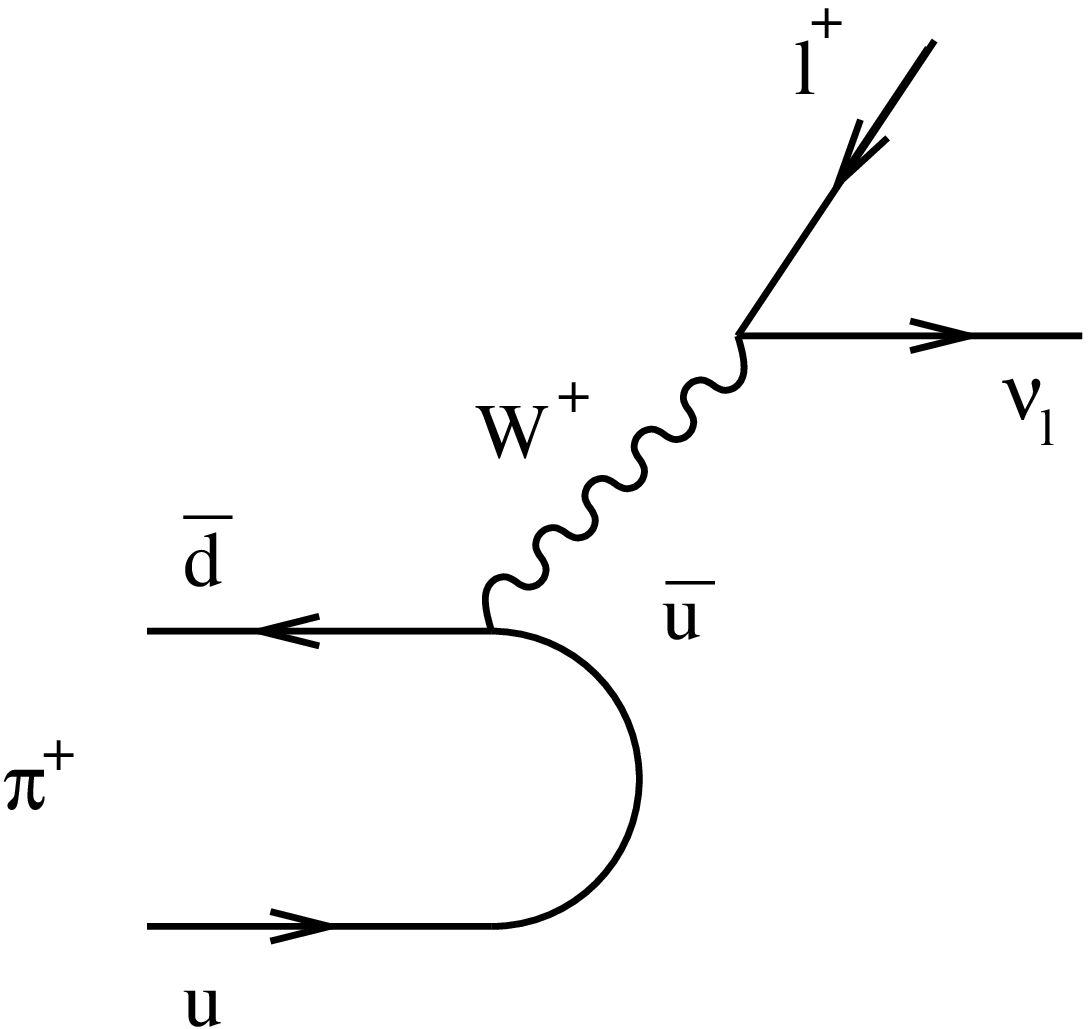,width=8cm}

\vspace{1cm}
\caption{}
\label{fig:tmpsn1}
\end{center}
\end{figure}

Let us consider $\pi^+ \to \mu^+ \nu_{\mu}$ decay (Fig.~\ref{fig:tmpsn1}).
The amplitude of the process has a kind:
\begin{eqnarray}
\label{eq:M}
M=\frac{G_F}{\sqrt{2}} V^*_{ud} \bar u_{\nu}\gamma^{\alpha}(1-\gamma_5)
v_{\mu} <0\mid J_{\alpha} \mid \pi^+>, 
\end{eqnarray}
where the phenomenological matrix element of a pseudovector current can be
submitted as
\begin{eqnarray}
\label{eq:Mt}
<0\mid J_{\alpha} \mid \pi^+> = \varphi_{\pi} f_{\pi} p_{\alpha}. 
\end{eqnarray}
Here $f_{\pi}$ is a constant of a charged pion, $f_{\pi}=0.131 GeV$ 
\cite{Eidelman:2004wy}, $\varphi_{\pi}$ is a wave
function of a pion, and $p_{\alpha}$ is its four-momentum.

As well as in paper \cite{Efrosinin:2005mr} we record a pseudovector current as a current
of constituent quarks
\begin{eqnarray}
\label{eq:Mt1}
<0 \mid J_{\alpha} \mid \pi^+> = \bar v_{\bar d}\gamma_{\alpha}\gamma_5
u_u F_{\pi}, 
\end{eqnarray}
here $\bar v_{\bar d},u_u$ are four-spinors of constituent quarks, $F_{\pi}$ is
parameter of an annihilation in vacuum, having dimension of mass. $F_{\pi}$
corresponds to an annihlation of a virtual pair of $u\bar u$. And the degree of
this process (departure from a mass surface) corresponds to energy of a pair
lepton-antilepton or mass of a pion.

In the rest frame of a pion from (\ref{eq:Mt1}) by simple calculation is
received
\begin{eqnarray}
\label{eq:Mt2}
<0 \mid J_{\alpha} \mid \pi^+> = F_{\pi} 2 m_u, 
\end{eqnarray}
where $m_u$ is mass of a constituent $u(d)$ quark, as well as in the paper
\cite{Efrosinin:2005mr} and also in the papers \cite{Gershtein:1976aq,
Khlopov:1978id}. From expressions (\ref{eq:Mt}) and (\ref{eq:Mt2}) is received
\begin{eqnarray}
\label{eq:Mte}
f_{\pi} m_{\pi}=F_{\pi} 2m_u, 
\end{eqnarray}
$f_{\pi}=0.131~GeV$ \cite{Eidelman:2004wy}.

Similarly for $K^+ \to \mu^+ \nu_{\mu}$ decay we have
\begin{eqnarray}
\label{eq:MtK}
f_K m_K=F_K 2\sqrt{m_u m_s}, 
\end{eqnarray}
$f_K=0.160~GeV$ \cite{Eidelman:2004wy}.

And also for $D^+_s \to \mu^+ \nu_{\mu}$ decay is
\begin{eqnarray}
\label{eq:MtD}
f_{D^+_s} m_{D^+_s}=F_{D^+_s} 2\sqrt{m_c m_s}, 
\end{eqnarray}
$f_{D^+_s} \simeq 0.288~GeV$ \cite{Eidelman:2004wy}.
.

Using (\ref{eq:Mte},\ref{eq:MtK},\ref{eq:MtD}) and also masses of constituent
quarks
\begin{eqnarray*}
\label{eq:MtM}
m_u &=& 0.305~GeV~  \cite{Efrosinin:2005mr},\\
m_s &=& 0.487~GeV~  \cite{Efrosinin:2005mr},\\
m_c &=& 1.400~GeV~  \cite{Eidelman:2004wy}  
\end{eqnarray*}
is received
\begin{eqnarray}
\label{eq:MtF}
F_{\pi^+} &=& \frac{f_{\pi} m_{\pi}}{2 m_u} = 0.030~GeV,\nonumber\\
F_{K^+} &=& \frac{f_K m_K}{2 \sqrt{m_u m_s}} = 0.103~GeV,\nonumber\\
F_{D^+_s} &=& \frac{f_{D^+_s} m_{D^+_s}}{2 \sqrt{m_c m_s}} = 0.343~GeV. 
\end{eqnarray}

We are interested by a prediction for a constant of $D^+ \to \mu^+\nu$ decay.
In paper \cite{Eidelman:2004wy} for parameter $f_{D^+}$ the high bound is given
only 
\begin{eqnarray}
\label{eq:Eid}
f_{D^+} < 310~MeV.
\end{eqnarray}

\begin{figure}[ht]
\begin{center}
\epsfig{file=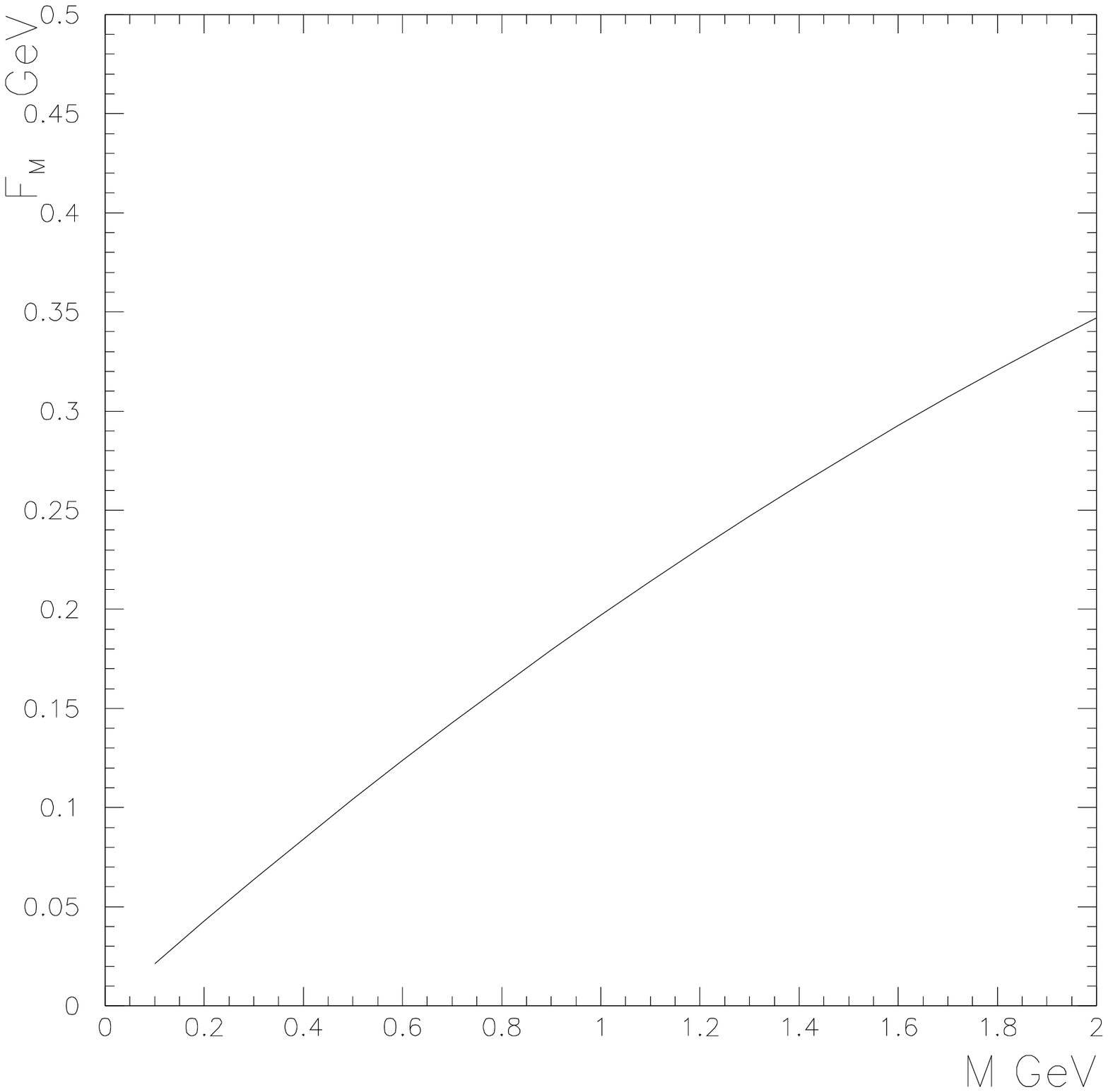,width=12cm}

\vspace{1cm}
\caption{}
\label{fig:tmpsu11}
\end{center}
\end{figure}

We have suspected that parameter $F_{M^+}$ depends only on mass of a meson.
We search for this relation as 
\begin{eqnarray}
\label{eq:fom}
F_{M^+} = aM^2+bM+c.
\end{eqnarray}
With usage of (\ref{eq:MtF}) we discover
\begin{eqnarray}
\label{eq:fom1}
F_{M^+} = -0.0238M^2+0.2213M-0.0005~(GeV).
\end{eqnarray}
The behavior of parameter $F_{M^+}$ depending on mass is submitted in 
 (Fig.~\ref{fig:tmpsu11}).
 
Let us remark that a matrix element in (\ref{eq:Mt1}) is half outside of a
mass surface. And is approximated as on mass surface. The effect of departure
from a mass surface is actuated in parameter of an annihilation $F_{M^+}$.
The difference of these effects for pairs $u\bar u$ and $s\bar s$ is
supposed miner.  

From the formula (\ref{eq:fom1}) is received
\begin{eqnarray}
\label{eq:fo1}
F_{D^+} = 0.33~ GeV.
\end{eqnarray}
Then from expression
\begin{eqnarray}
\label{eq:fom2}
F_{D^+}m_{G^+} = F_{D^+} 2\sqrt{m_cm_u}
\end{eqnarray}
we have
\begin{eqnarray}
\label{eq:fom3}
f_{D^+} = 0.23~ GeV,
\end{eqnarray}
that does not contradict (\ref{eq:Eid}).

The obtaining of the new data of constants of decay of pseudoscalar mesons will
help to explain a phisical essens of parameter of an annihilation $F_{M^+}$.
The introducing of this parameter encourages division of series stages weak and
strong interactions in decaies of pseudoscalar mesons and refinement of
mechanism of these decaies.



\newpage
\clearpage

\newpage
\clearpage

\begin{center}
Figure captions
\end{center}

Fig.~1. The diagram  of $\pi^+ \to \mu^+ \nu_{\mu}$ decay
        on a quark level.

Fig.~2. Relation of parameter of an annihilation $F_{M^+}$ to mass of
        a pseudoscalar meson.


\end{document}